# Coupled Schrödinger equations as a model of interchain excitation transport in the DNA

Margarita Kovaleva[1,2,*], Leonid Manevitch[1]

[1] N.N. Semenov Federal Research Center of Chemical Physics Russian Academy of Sciences, 4 Kosygin st., Moscow 119991, Russia

[2] Faculty of Physics, HSE University, 21/4 Staraya Basmannaya st., Moscow 105066, Russia

* corresponding author: makovaleva@chph.ras.ru

**Abstract:** In our report we consider two weakly coupled Schrödinger equations as a model of the interchain energy transport in the DNA double-helix. We use the reduction of the Yakushevich-type model considering the torsional dynamics of the DNA. In the previous works only small amplitude excitations and stationary dynamics were investigated, while we focus on the non-stationary dynamics of the double-helix. We consider the system as a model of two weakly interacting DNA strands. Supposing that initially only one of the chains is excited in form of breather we demonstrate the existence of invariant which allows to reduce the order of the problem and consider the system of the phase plane. The analysis provided demonstrates analytical tool for prediction of the periodic interchain excitation transitions of its localization on one of the chains. The technique also takes into account the spreading of the excitations with time.

## 1. Introduction

Dynamical properties of DNA attract attention of researchers from different fields of science during decades[1]. Molecule of DNA is an extraordinary complex system. However, some of its features can be investigated at the fundamental level. The macromolecule's structural transformations define its dynamical properties. Many dynamical aspects of the DNA functioning which are related only with the weak electrostatic bonds can be conventionally analyzed withing a framework of dynamical modelling. The dynamics of DNA molecule can be important in different biological processes, such as replication, conformation transitions, transcription and formation of the higher-order structures[2]. Some of the processes can be also important in applications.

There are many different models of the DNA dynamics. For the molecular dynamical modeling full-models are used which take into account all the potentials of interaction, many also consider the solvent and ions actions[3,4]. Such techniques allow deeper understanding of processes' the mechanisms; however, they are very time-consuming and cannot be used for analysis of long fragments of the DNA for long periods of time. To study the dynamics on the higher time-scales, it is necessary to use reduced model taking into account only some structural elements and dynamical features to obtain more general results[4,5,6]. There are many attempts to reduce the DNA model with use of the coarse graining. Some of them were proved to be quite reasonable for description of different processes in the double helix[7], including transitions between the A-form and B-form[8]. Thus, to make any semi-analytical predictions of the dynamics of the DNA double-helix more reductions are necessary.

Different dynamical processes in DNA are associated with different models. Replication and denaturation processes imply separation between the two DNA strands. Therefore, they are better considered in context of the models with the radial evolution. The pioneering model of the type by Bishop and Peirard[9] was later extended by Dauxois and coauthors[10], and by Barbi with colleagues[11]. Another type of models is connected with the transcription process, which is defined mostly by the torsional (rotational) excitations. For the latter case the first developed model was proposed by Yakushevich[12]. The model considers only the torsional dynamics of the double-helix together with the bases connecting the strands. This model was later modified to the 'composite' model by Cadoni et al[13].

Our main attention is paid to the dynamics of the DNA molecule connected with the replication process. We would like to demonstrate study of unusual regimes in the locally excited DNA. Therefore, we consider the thoroughly proved model of the Yakushevich type [15]. We are interested mostly in the inter-chain breathers transport study. Such a work was first initiated by Kovaleva et al.[16], but only phenomenologically, and the spreading of the excitations was not taken into account, as well as the phase evolution of the reduced system was not cleared. In our paper we use regular multiple-scales procedure to reduce the Yakushevich-type model[17] to the system of two coupled Schrödinger equations.

Nonlinear Schrödinger equation is a universal tool used for description of many nonlinear systems and objects in different field of science, such as optics, physics of plasmas, solid-state physics etc. Being classical this equation is studied in details[18-23]. However, the interchain excitations exchange in the system of coupled Schrödinger equations is not well studied. Some reductions and conclusions were presented in [14]. We present the detailed study of the interchain breathers exchange on the simplified model of the DNA.

We consider the breather, corresponding to localized excitation of one of the strands of DNA. Applying the collective coordinates technique, we suppose that the form of the of excitation is not changed via inter-chain interactions. We reveal the new invariant of the reduced-order system, is allows us to represent the evolution of the system on the phase plane. The analysis gives the value of system's parameters for the dynamic threshold, when transition from interchain excitation exchange to excitation localization occurs. We prove our analysis by results of the numerical modelling of the initial model.

## 2. Model formulation

Let us consider the B form of a DNA molecule; the double-helix is considered as two coupled chains of equivalent pendula. The Hamilton function of the double-strand is described by the angular displacements of the n-th base at the first and the second chains correspondingly:

$$H = \sum_n \left\{ \frac{1}{2} I_{n,1} \dot{\psi}_{n,1}^2 + \frac{1}{2} I_{n,2} \dot{\psi}_{n,2}^2 + \varepsilon_{n,1} \sin^2 \frac{\psi_{n+1,1} - \psi_{n,1}}{2} + \varepsilon_{n,2} \sin^2 \frac{\psi_{n+1,2} - \psi_{n,2}}{2} + V_{\alpha,\beta}(\psi_{n,1}, \psi_{n,2}) \right\}$$

$$V_{\alpha,\beta}(\psi_{n,1}, \psi_{n,2}) = C_1(1 - \cos \psi_{n,1}) + C_2(1 - \cos \psi_{n,2}) - C_{12}(1 - \cos(\psi_{n,1} - \psi_{n,1})),$$

(1)

where the parameters are defined as follows:

$$C_1 = K_{\alpha\beta} r_\alpha (r_\alpha + r_\beta), \quad C_2 = K_{\alpha\beta} r_\beta (r_\alpha + r_\beta), \quad C_{12} = \frac{1}{4} K_{\alpha\beta} \left(1 - \frac{\omega_{\alpha\beta 2}}{\omega_{\alpha\beta 1}}\right) (r_\alpha + r_\beta)^2,$$

$\omega_{\alpha\beta 1}, \omega_{\alpha\beta 2}$ are the frequencies of the rotational vibrations of the same and the opposite directions respectively, $K_{\alpha\beta}$ corresponds to the stiffness of the base-pair interaction. The interaction along the chains is accounted in the nearest-neighbor approximation. Two first terms of the Hamiltonian represent the kinetic energy of the n-th base pair, $I_{n,1}, I_{n,2}$ are the inertia moments of the n-th base of the first and the second chains respectively. The third and the forth terms describe the interaction of the neighboring base pairs along each of the macromolecule chains. Parameter $\varepsilon_{n,1} = \varepsilon_{n,2} = \varepsilon_{al}$ characterizes energy of interaction of the n-th base with the (n+1)-th base of the *i*-th (*i*=1,2) chain. The last term of the Hamiltonian represents energy of interaction between the coupled bases of different chains.

The model itself, its spectra and the possible solutions were discussed in previous works[17,25]. However, the localized excitations wondering between the chains were not studied in details.

## 3. Breathers and their inter-chain transport

The model (1) admits different types of the localized solutions, including solitons and breathers. In the numerical studies of the coupled system (1) we have observed the breathers wondering between the two chains or breathers' localization on one chain depending on the value of the coupling parameter between the two chains. To proceed with the further observations of the phenomenon we make some assumptions concerning the properties of the excitation. In the limit of small amplitudes and the long-wave approximation the equations of motion can be transformed to quasi-continuum and two coupled nonlinear Schrödinger equations can be obtained (see Appendix for details):

which can be transformed to a simple form:

$$i\frac{d\tilde{\varphi}_1}{d\tau} = \frac{d^2\tilde{\varphi}_1}{dx^2} + |\tilde{\varphi}_1|^2 \tilde{\varphi}_1 + \mu(\tilde{\varphi}_1 - \tilde{\varphi}_2),$$
$$\frac{d\tilde{\varphi}_2}{d\tau} = \frac{d^2\tilde{\varphi}_2}{dx^2} + |\tilde{\varphi}_2|^2 \tilde{\varphi}_2 - \mu(\tilde{\varphi}_2 - \tilde{\varphi}_1). \quad (5)$$

In the uncoupled system, when $\mu = 0$ the breather solution can be found as follows:

$$\tilde{\varphi}_1 = A_1 \text{sech}(a_1(x - V_1\tau))\exp(i(x - V_1\tau)\xi_1 + i\sigma_1);$$
$$\tilde{\varphi}_2 = A_2 \text{sech}(a_2(x - V_2\tau))\exp(i(x - V_2\tau)\xi_2 + i\sigma_2); \quad (6)$$

To study the possibility of the breather wondering between the chains or localization on one chain we proceed to the collective coordinates technique [26]. We suppose that the breather wondering between the chains is slow due to weak coupling between the chains. This allows us to consider the solution in the form (5) but all the parameters being super-slow time-dependent (the new time-scale is defined as $\tau_1$). The Lagrange function of the system looks as follows:

$$L = \sum_n \left[ \frac{i}{2}\left(\frac{\partial \tilde{\varphi}_n}{\partial \tau_1}\tilde{\varphi}_n^* - \frac{\partial \tilde{\varphi}_n^*}{\partial \tau_1}\tilde{\varphi}_n\right) - \left|\frac{\partial \tilde{\varphi}_n}{\partial x}\right|^2 + |\tilde{\varphi}_n|^4 \right] + \mu|\tilde{\varphi}_1 - \tilde{\varphi}_2|^2, n = 1, 2$$

All the parameters of the solution (4) serve as collective coordinates of the system. Performing the first variation of the effective Lagrangian with respect to the collective coordinates, we show that the solution (4) can exist in the coupled system in case $V_1 = V_2$, and it admits an invariant:

$$N = \frac{\xi}{a}(A_1^2 + A_2^2),$$

where $\xi_1 = \xi_2 = \xi$, $a_1 = a_2 = a$.

Evidently enough, we can suppose that $\xi = const$, then we obtain the integral in the following form:

$$N_0 = \frac{1}{a}\left(A_1^2 + A_2^2\right), \tag{7}$$

Such invariant allows consider weak spreading of the excitation and describe dynamics with dispersion taken into account. When the amplitude of the excitation is decreased, the width of the solution grows.

If we suppose that the dispersion is weak enough to allow interchain breathers exchange, then we can suppose that the spreading with time, i.e. neglect the $\frac{da}{d\tau_1}$ in the leading approximation. Then the integral of motion takes the form of the norm,

$$N_1 = A_1^2 + A_2^2.$$

Such an integral allows to introduce angular representation:

$$\tilde{\varphi}_1 = N_1 \cos(\theta(\tau_1)) e^{i\delta_1} \varphi(x, \tau);$$
$$\tilde{\varphi}_2 = N_1 \sin(\theta(\tau_1)) e^{i\delta_2} \varphi(x, \tau);$$
$$\Delta = \delta_1 - \delta_2, \tag{8}$$

where $\varphi(x, \tau)$ – is soliton solution of the isolated chain with amplitude equal to unity.

Now the equations of motion can be transformed to the following form:

$$\frac{d\theta}{d\tau_1} = \sin(\Delta),$$
$$\frac{d\Delta}{d\tau_1} \sin(2\theta) = k \sin(2\theta) + \cos(\Delta)\cos(2\theta), \tag{9}$$

The evolution of the system can be represented on the phase plane.

We start the analysis of (9) by seeking of the system's fixed points, which are crucial for the proper understanding of the whole mechanism triggering the inter-chain energy transfer. The first fixed point $(\Delta_1, \theta_1)$ which corresponds to the in-phase periodic motion of the full system with an equal energy distribution between the coupled chains, is derived from (9):

$$\Delta_1 = 0, \theta_1 = \frac{\pi}{4}.$$

There are three additional fixed points which correspond to the out-of-phase periodic motion of the full model $(\Delta_2, \theta_2)$. The stationary solution of the modulated system corresponding to the out-of-phase motion of the two oscillators while energy is being equally distributed between the two, looks as follows:

$$\Delta_2 = \pi, \theta_2 = \frac{\pi}{4}.$$

For simplicity we consider only coordinates in the range $0 \leq \theta_k \leq \frac{\pi}{2}$, $-\pi < \Delta_k \leq \pi$ due to the periodicity of the angular coordinates $(\theta, \Delta)$.

In terms of the super-slow flow model, the regime of complete energy exchanges between the chains corresponds to a special orbit that passes through $\theta = 0$ and reaches the value of $\theta = \pi/2$. This special type of trajectory is referred to in the literature as a *Limiting Phase Trajectory* (LPT). The concept of LPTs was introduced first in [27] and it appears to be very useful methodology for an adequate description and understanding of nonstationary resonant processes emerging in the various oscillatory models (see also [28]). The LPT is a nonstationary oscillatory regime, exhibiting a complete energy exchange between the coupled systems. The initial conditions corresponding to the LPT are $\theta = \Delta = 0$ which can also be interpreted as the complete initial energy localization on the first chain while the second one is at the rest.

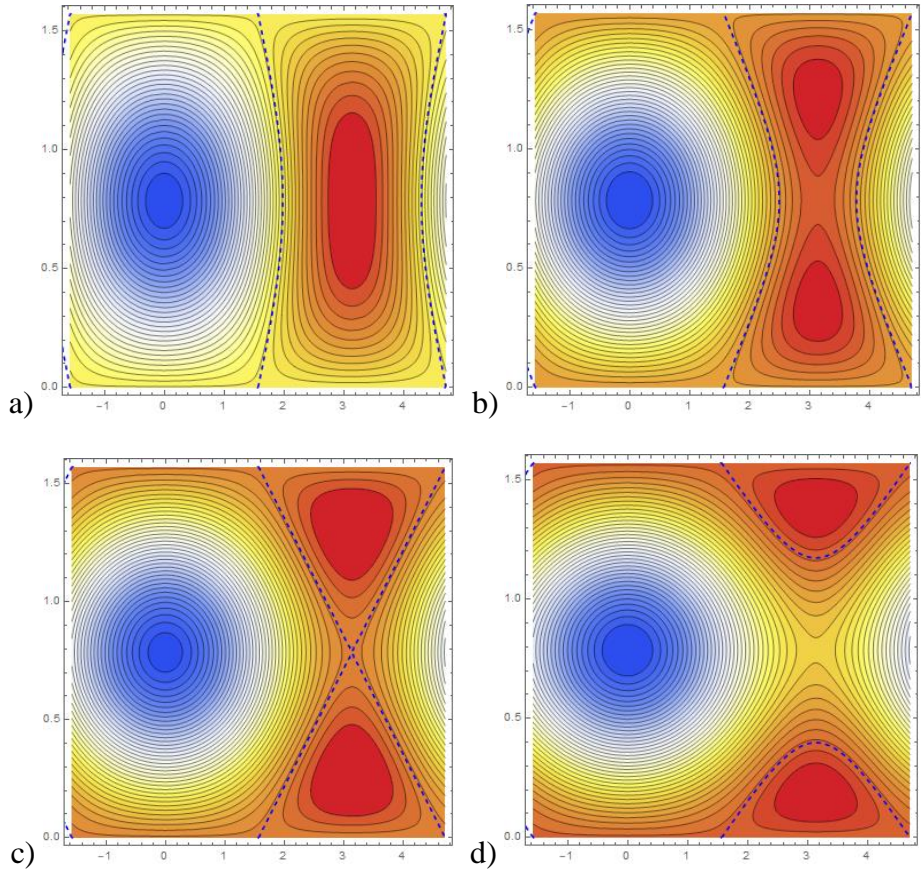

**Figure 1.** Phase plane of (9) corresponding to different values of coupling parameter: a) k = 0.2; b) k = 0.4; c) k = 0.5; d) k = 0.7. Blue dashed line denotes the LPT.

If the coupling parameter $k$ is low, there are only two stationary point on the phase plane, the LPT passes both the $\theta = 0$ and $\theta = \pi/2$ points when starting from one of them. If the coupling increases the anti-phase stationary point loses stability, two new stationary point are born, they are separated by a separatrix passing throw the saddle point. However, below the critical value $k_{cr}$=0.5 the LPT does not change its form. When the coupling parameter increases the new bifurcation occurs, the topological transformation of the phase space occurs, the LPT collides with the separatrix, and becomes localized around one of the newborn stationary points. This corresponds to localization of the excitation on one of the chains when the initial conditions correspond to excitation of one of the chains in the system.

## 4. Numerical verification of the results

To prove the results obtained in asymptotic analysis we look at the evolution of the system (1) where initial conditions correspond to excitation in form of breather placed only of one chain. According to the results discussed above we should observe localization of the breather on one of the chains or wandering of the breathers between the chains in the super-slow time-scale. We demonstrate at Figure 2 and Figure3 that prediction works quite well for the defined parameter range, corresponding to weak coupling and small amplitude excitations limit. As expected, the localized excitations spread, and the width of the signal increases, while the amplitude of excitation in decreased. Such kind of evolution is evidently supported by the invariant (7). We represent the evolution also as time space diagram (see Fig.4). Here we also see the realization of the interchain transport of the excitation. The profiles at different time-steps show the spread of the energy at the oscillating background, moving from the breather towards the ends of each chain However, until the amplitude of the background excitation is much lower than the amplitude of the breather solution, our analysis is still relevant.

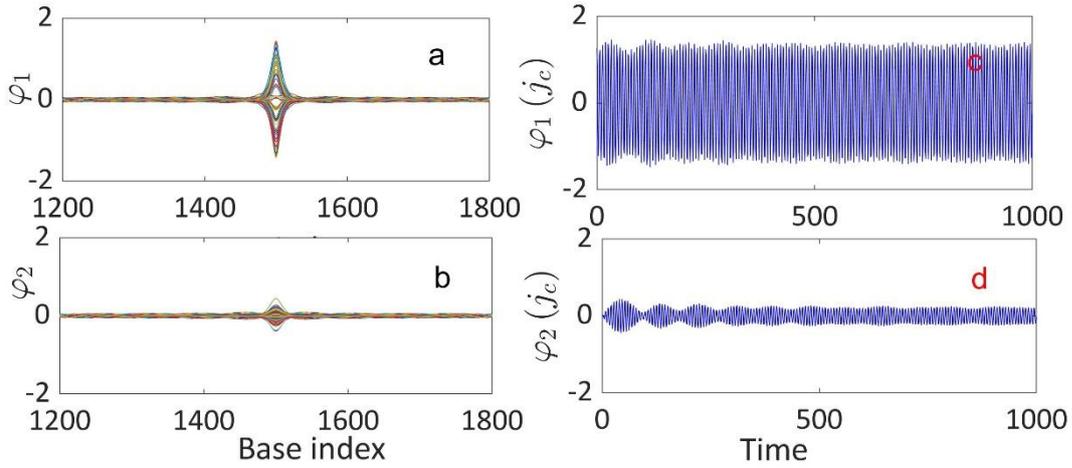

**Figure 1.** Breather profile and time evolution of the middle element corresponding to the controlling parameter $k = 0.5$, localization of the excitation: a) and b) profile snapshots in different moments of time; c) and d) time-evolution of the middle-element of the chain. All the simulations are provided for initial model (1).

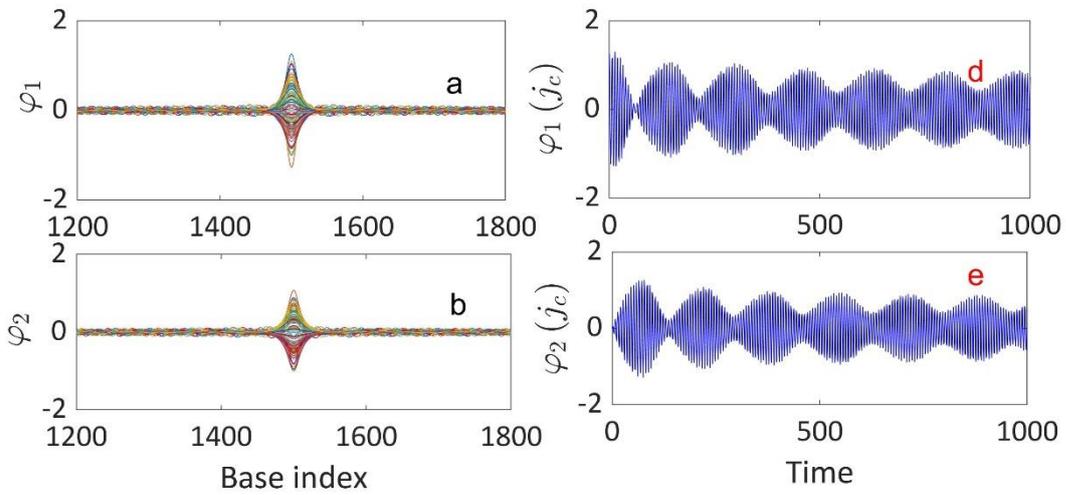

**Figure 2.** Breather profile and time evolution of the middle element corresponding to the controlling parameter $k = 0.2$, excitation exchange between the two chains: a) and b) profile snapshots in different moments of time; c) and d) time-evolution of the middle-element of the chain. All the simulations are provided for initial model (1).

**Figure 3.**

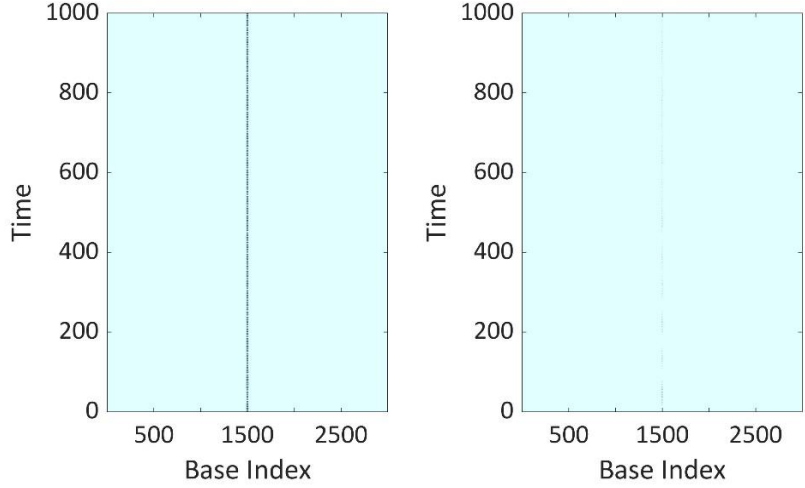

a)

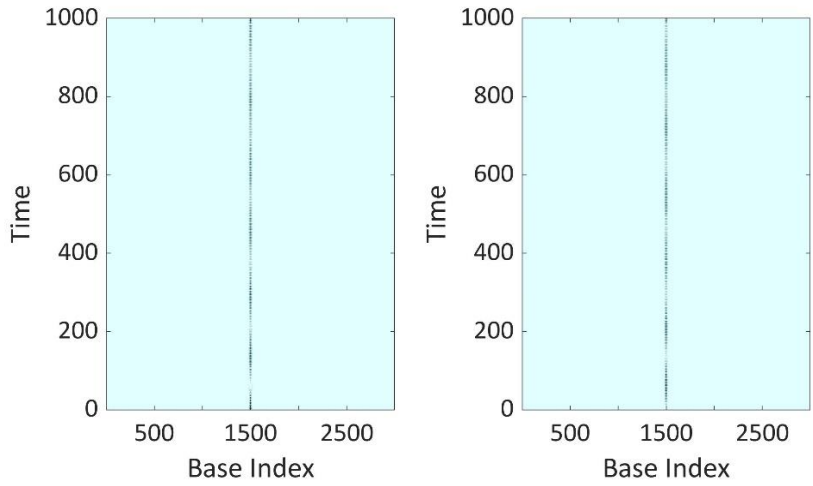

b)

**Figure 4.** Breather time-space evolution corresponding to the different controlling parameter values: a) $k = 0.2$, b) $k = 0.5$. Left panes show the evolution of the breather at the first chain, right panes demonstrate the evolution of the breather at the second chain. All the simulations are provided for initial model (1).

## 5. Conclusions

In current work we analyzed the process of breather excitation wondering between the chains of the double-helix of DNA in the low-amplitudes limit. We reduced the system to weakly coupled nonlinear Schrödinger equations. The effective Lagrangian and collective coordinates technique allow to reveal the invariant of the system in the regime of breathers' exchange between the two chains. We support our analysis by the numerical simulation of the initial system. The anti-phase and in-phase breathers' evolution was already reported in previous studies, however, the interchain transport was not considered in details, it was only reported phenomenologically. We proved with our analysis the existence of the regime of intensive excitation exchange and its transition to localization. With this study we supported the idea that even in the case of the spreading solutions in long chains some analytical predictions and reduced-order analysis can be provided. We believe that the type of the excitations inter-chain transport can be relevant in the others weakly non-linear systems described by the coupled Schrödinger equations.

**Acknowledgments**

This work was supported by Russian Science Foundation according to the research project no. 16-13-10302

**Appendix**

Starting the analysis with the Hamiltonian (1) we suppose that the amplitudes are small and proceed to the long-wave approximation:

$$\sin(\psi_{n+1,i} - \psi_{n,i}) - \sin(\psi_{n,i} - \psi_{n-1,i}) \approx \psi_{n+1,i} - \psi_{n,i} - (\psi_{n,i} - \psi_{n-1,i}) \approx h^2 \frac{\partial^2 \psi_{n,i}}{\partial z^2}, \qquad (1A)$$

Where $h$ is a typical distance between the bases, space variable $\chi = \sqrt{\varepsilon} z = \sqrt{\varepsilon} nh$, $\varepsilon$ is a small parameter characterizing relation between the distance $h$ and the wavelength of the excitation. Introducing the dimensionless time variable $\tau = \omega_0 t$, where $\omega_0 = \sqrt{K_{\alpha\beta} r_\alpha (r_\alpha + r_\beta) \frac{1}{I_2}}$, we obtain a system of nonlinear PDEs:

$$\ddot{\psi}_1 - \sqrt{\varepsilon} c_1^2 \frac{\partial^2 \psi_1}{\partial \chi^2} + \gamma_1 \sin \psi_1 - \eta_1 \sin(\psi_1 - \psi_2) = 0,$$
$$\ddot{\psi}_2 - \sqrt{\varepsilon} c_2^2 \frac{\partial^2 \psi_2}{\partial \chi^2} + \gamma_2 \sin \psi_2 - \eta_2 \sin(\psi_2 - \psi_1) = 0, \qquad (2A)$$

where $c_i^2 = \frac{\varepsilon_{n,i}}{2 I_i \omega_0^2}, \gamma_i = \frac{C_i}{I_i \omega_0^2}, \eta_i = \frac{C_{12}}{I_i \omega_0^2}$, the values of coefficients in both equations we suppose to be close to each other. The variable transformation $u = \sqrt{\varepsilon} \psi_1, v = \sqrt{\varepsilon} \psi_2$ and the expansion of the trigonometric functions in the small-amplitude limit yields:

$$u_{\tau\tau} - \varepsilon c_1^2 u_{\chi\chi} + \gamma_1 u - \frac{1}{6} \gamma_1 u^3 - \eta_1 (u - v) + \frac{1}{6} \eta_1 (u - v)^3 = 0,$$
$$v_{\tau\tau} - \varepsilon c_2^2 v_{\chi\chi} + \gamma_2 u - \frac{1}{6} \gamma_2 v^3 + \eta_2 (u - v) - \frac{1}{6} \eta_2 (u - v)^3 = 0, \qquad (3A)$$

To provide the asymptotic analysis we apply the complexification of the variables:

$$\dot{u} = \frac{1}{2}(\Psi_1 + \Psi_1^*); u = -\frac{1}{2i}(\Psi_1 - \Psi_1^*); \quad \dot{v} = \frac{1}{2}(\Psi_2 + \Psi_2^*); v = -\frac{1}{2i}(\Psi_2 - \Psi_2^*);$$

Separating "fast" $\tau_0 = t$ and "slow" $\tau_1 = \varepsilon \tau_0$ time-scales we use the two-scale procedure:

$$\Psi_j(\tau_0, \tau_1) = \Psi_j^0(\tau_0, \tau_1) + \varepsilon \Psi_j^1(\tau_0, \tau_1), j = 1, 2.; \quad \frac{d}{d\tau} = \frac{\partial}{\partial \tau_0} + \varepsilon \frac{\partial}{\partial \tau_1}$$ to obtain the system in the leading asymptotic approximation ($O(1)$):

$$\frac{d\Psi_j^0}{dt} - i\Psi_j^0 = 0, j = 1, 2.$$

Solution gives the following form for the variables:

$$\Psi_j^0(\tau_0, \tau_1) = \varphi_j(\tau_1) \exp(i\tau_0), j = 1, 2.$$

In the next order of the expansion we obtain the system of PDEs:

$$i\frac{d\tilde{\varphi}_1}{d\tau_1} = \frac{d^2 \tilde{\varphi}_1}{d\chi^2} + \alpha_1 |\tilde{\varphi}_1|^2 \tilde{\varphi}_1 + \mu_1(\tilde{\varphi}_1 - \tilde{\varphi}_2),$$

$$\frac{d\tilde{\varphi}_2}{d\tau_1} = \frac{d^2 \tilde{\varphi}_2}{d\chi^2} + \alpha_2 |\tilde{\varphi}_2|^2 \tilde{\varphi}_2 - \mu_2(\tilde{\varphi}_2 - \tilde{\varphi}_1),$$
(4A)

Using this system we can easily obtain the system of equations (5).

## References

[1] Calladine, C., Drew, H., Luisi, B., and Travers, A.. Understanding DNA. Academic press, London, 2004

[2] Saenger, W. Principles of Nucleic Acid Structure. Springer Verlag, New York, 1984

[3] Beveridge, D.L., Cheatham, T.E., Mezei, M.: The ABCs of molecular dynamics simulations on B-DNA. *J. Biosci.* 37 379–397(2012).

[4] Beveridge, D.L. and McConnell K.J., Nucleic acids: theory and computer simulation, Y2K, *Curr. Opin. Struct. Biol*. 2 10, 182-196(2000).

[5] Yakushevich L.V., Nonlinear DNA dynamics: hierarchy of the models. *Physica D Nonlinear Phenomena* 79,77-86(1994).

[6] Cheatham III, T. E., Molecular Modeling and Atomistic Simulation of Nucleic Acids Elsevier, Amsterdam, Vol.1, Chap. 6, pp. 75–89 (2005).

[7] Kovaleva, N.A., Koroleva (Kikot), I.P., Mazo, M.A. et al.The "sugar" coarse-grained DNA model *J Mol Model* 23, 66(2017).

[8] Kovaleva, N.A., Zubova, E.A. MD simulation of the transitions between B-DNA and A-DNA in the framework of a coarse-grained model *Doklady Physical Chemistry* 475, 119–121(2017).

[9] Peirard M., Bishop A.R. Statistical Mechaics of a non-linear model for DNA denaturation *Phys Rev Lett*, 62 (23), 2755-2758 (1989).

[10] Dauxois T., Peirard M. and Bishop A.R. Thermodynamics of a nonlinear model for DNA denaturation *Phys D*, 66 (1-2): 35-42(1993).

[11]  Barbi M., Cocco S., and Peirard M. Helicoidal model for DNA opening *Phys Lett A*, 253, 358-369 (1999)

[12]  Yakushevich L.V.  Nonlinear DNA dynamic: A new model. *Phys Lett A*, 136, 413-417(1989)

[13]  Cadoni M., De Leo R., and Gaeta G.  A composite model for DNA torsion dynamics *Phys Rev. E*, 75, 021919(2007)

[14]  Dauxois T., Peyrard M., A nonlinear model for DNA melting in Nonlinear excitation in biomolecules / Ed. by Peyrard. Springer, 1995. 127-136

[15]  Yakushevich, L.V., Nonlinear Physics of DNA, Willey, Weinheim, 2004.

[16]  Kovaleva N.A. Nonlinear torsional dynamics of the two-strand discrete DNA model, PhD thesis, 2005 (russian) https://dlib.rsl.ru/viewer/01003286357

[17]  Yakushevich, L.V., Savin, A.V., and Manevitch, L.I., Nonlinear dynamics of topological solitons in DNA *Phys. Rev. E*. 66, 016614(2002).

[18]  Ershov, S. N., Vaagen , J. S., and Zhukov M. V.  Modified variable phase method for the solution of coupled radial Schrödinger equations *Phys. Rev. C* 84, 064308 (2011)

[19]  Wadati, M. Iizuka T., Hisakado M. A Coupled Nonlinear Schrödinger Equation and Optical Solitons *J of the Physical Society of Japan*, 61, 7, 2241-22459(1992)

[20]  Abrarov, R.M., Christiansen, P.L., Darmanyan, S.A., Scott A.C., Soerensen, M.P. Soliton propagation in three coupled nonlinear Schrödinger equations *Physics Letters A* 171, 5–6, 298-302(1992).

[21] Todorov, M.D. Systems of coupled nonlinear Schrödinger equations. Vector Schrödinger equation / Nonlinear Waves Theory, computer simulation, experiment Morgan & Claypool Publishers, 2018, pp 2-1 - 2-118.

[22] Abdelrahman M.A.E, Hassan S. Z.  Nonlinear Schr¨odinger Equation: Symmetries and Exact Solutions *Modern Physics Letters B* 34, 2050078 (2020).

[23] Borhanifar A., Abazari R. Numerical study of nonlinear Schrödinger and coupled Schrödinger equations by differential transformation method *Optics Comm* 283, 10, 2026-2031(2010).

[24] Gendelman, O.V., Manevitch, L.I., Tractable Models of Solid Mechanics: Formulation, Analysis and Interpretation, Springer-Verlag, Berlin 2011.

[25] Kovaleva, N.A., Manevich, L.I., Musienko, A.I. et al. Low-frequency localized oscillations of the DNA double strand. Polym. Sci. Ser. A 51, 833–847 (2009)

[26] Scott, A. Nonlinear Science. Emergence and dynamics of coherent structures, Oxford University Press, 2005

[27]  Manevitch L. I., New approach to beating phenomenon in coupled nonlinear oscillatory chains. Arch Appl Mech 77, 301(2007).

[28]  Manevitch L.I., Kovaleva A., Smirnov V., Starosvetsky Yu.  Nonstationary Resonant Dynamics of Oscillatory Chains and Nanostructures. Springer; 2017